# Origin of the backstreaming ions in a young Hot Flow Anomaly.

O.L.Vaisberg, S.D.Shuvalov, A.Yu.Shestakov, Y.M.Golubeva

**Abstract**

We analyze an event in front of the bow shock observed by Cluster spacecraft on 22.02.2006. This event has many attributes of Hot Flow Anomaly at early stage of development including strong upstream beam and disturbed magnetic profile with increased magnetic field at one or two sides as observed by 4 Cluster spacecraft. The angle between the magnetic field vectors at two sides of the current sheet was ~ 10°. The minimum magnetic field magnitude within HFA was ~ 1 nT. The shock at two sides of the HFA was quasi-perpendicular. Upstream beam was observed on the leading side of the HFA. Parameters and velocity distributions of solar wind ions and of upstream ions observed on C1 and C3 spacecraft are analyzed separately in order to trace their changes across the event. The goal of this analysis was to get more information about the source of upstream beam. The beam evolved from the start of its observation till the HFA encounter being initially energetic and nearly mono-energetic. Its mean energy continuously decreased and energy spectrum widened as HFA approached spacecraft. First observation of particular energy that diminished with approaching the HFA varied linearly with gyro-radius of ions. The energy spectra of integrated beam (for all observation) and the energy spectrum of the beam observed just in front of HFA are very similar to the magnetosheath ion energy spectrum observed after bow shock crossing at ~ 1 hour after observation of the HFA. Lowest energies in the beam were observed within HFA only. Highest density and pressure of upstream beam are found in the current sheet itself. We suggest that the upstream beam is the result of the magnetosheath ions leakage through the region of HFA crossing with the bow shock front.

## 1. Introduction

Hot Flow Anomalies were discovered in 1980[th] (Schwartz et al., 1985) and were explained as the results of interplanetary current sheets' interaction with the bow shock (Schwartz et al., 1988). Observations (Paschmann et al., 1988, Thomsen et al, 1988) and simulations (Burgess and Schwartz, 1988) indicated that HFAs develop due to interaction of the beam reflected from the bow shock with the solar wind in the vicinity of the discontinuity (Burgess, 1989). Two-stream distribution of ions excites the instability resulting in a strong flux heating and deceleration (Thomas and Brecht, 1988, Gary, 1991). Majority of these discontinuities are tangential discontinuities (Schwartz, 2000). Conditions of HFAs formation were determined in (Schwartz, 2000) and the list of conditions was extended in (Facsko, 2009).

There are few classifications of HFAs including separation on mature and young HFAs (Zhang, 2010) by ion velocity distribution relaxation inside HFA, and several types, distinguished by variation trend of the dynamic pressure during the event (Wang et al., 2012). Subsequently a few more classes of disturbances were found that mimic many characteristics of HFAs, including the foreshock cavities that were explained as the regions in which magnetic field is connected to the bow shock (Sibeck et al., 2002).

Foreshock bubbles (FB) that form as a result of the change in the interplanetary magnetic field direction associated with solar wind discontinuities and its interaction with the backstreaming ion beams in the foreshock. Unlike HFAs, foreshock bubbles are formed prior to the interaction of the discontinuity with the bow shock itself. FB were predicted in 2010 (Omidi, 2010) and later observed (Turner, 2013).

In this communication we are considering young HFA. The goal of this paper is to analyze characteristics of upstream beam that interact with the current sheet.

## 2. Observations
## 2.1. Selection of event

Event for analysis was chosen from the list of HFAs published in (Facsko et al. 2009). HFA was selected by criterion that all four spacecraft observed it at the early stage when the current sheet is easily identified (Figure 1). It is event #3 of the series of the active current layers observed on 22.02.2006 at ~ 01:08 UT (Figure 2). Characteristics of the solar wind beam and upstream beam were analyzed separately by division of the velocity space (see section 3 and Figure 3 for explanation) occupied by two beams (see explanation below). Ion parameters were calculated as moments of distribution in which phase space density was assigned to respective Cluster ($\theta$, $\varphi$, V) grid blocks. With 4-sec temporal resolution of CIS1/CODIF ion analyzer on C1 spacecraft upstream beam was documented in sufficient details for analysis. This beam was also recorded by CIS1/CODIF ion analyzer on C3 spacecraft with 12-sec resolution. The data of CIS2/HIA on C4 spacecraft were not used due to low counting rate at this time.

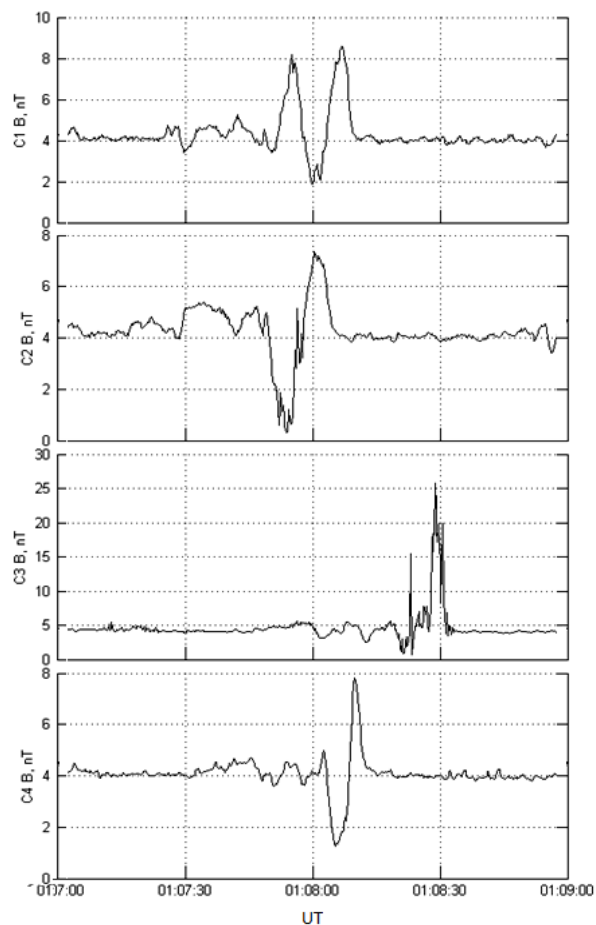

Figure 1. Magnetic field magnitudes measured by Cluster spacecraft during observation of HFA at 01:08 UT. Note the distorted current sheet profile, increased field regions at the sides of the current sheet, and the train of magnetic field variations at the forward side of the current sheet.

Increased magnetic fields at one or two sides were observed by every spacecraft. The structure of the current sheet is disturbed in the region of the minimum magnetic fields.

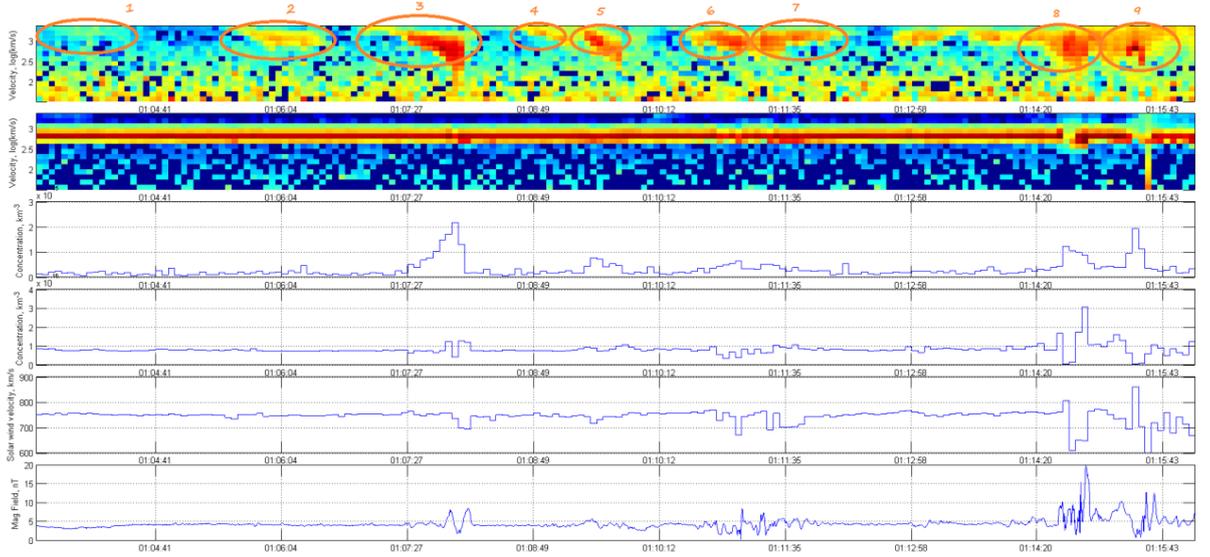

Figure 2. C1 spacecraft data. From top to bottom: upstream ion velocity-time spectrogram, solar wind velocity-time spectrogram, upstream beam density, solar wind density, solar wind velocity, and magnetic field magnitude.

Coordinates of 4 Cluster spacecraft are given in Table 1. The separation scale of the spacecraft is of order of 1 $R_E$ on average.

Table 1. GSE coordinates of Cluster spacecraft during observations shown in Figure 1.

| Distance, $R_E$ | Spacecraft | | | |
|---|---|---|---|---|
| | C1 | C2 | C3 | C4 |
| X | 10.665 | 10.669 | 9.832 | 10.912 |
| Y | - 2.400 | - 2.209 | - 3.019 | - 2.699 |
| Z | -11.026 | -10.796 | -11.873 | -11.862 |

## 2.2. HFA Environment

Solar wind velocity $V_{sw}$ and number density $n_{sw}$ calculated separately for the solar wind beam are: $V_{sw}$ ~ 750 km/s and $n_{sw}$ = 7.5 cm$^{-3}$. Average values of interplanetary magnetic field $\boldsymbol{B_1}$ and $\boldsymbol{B_2}$ at two sides of the current sheet were calculated from measurements on 4 spacecraft within time interval of less turbulent magnetic fields before and after current sheet crossing. Durations of intervals were about 30 s, averaged $\boldsymbol{B_1}$ equals to 4.13 (2.03; -3,29; 1,43) nT, averaged $\boldsymbol{B_2}$ equals to 3.94 (2.21; -2.77; 1.72) nT. The angles of magnetic field rotation at the current sheet were 12.8°±1.7° (C1), 7.5°±2.1° (C2), 6.8°±1.9° (C3), 6.9°±1.0° (C4). Errors of rotation angles' determination were calculated as standard error of the mean value (SEM), based on magnetic field variations during considered intervals. Error values were calculated for the interval of 3σ deviation from the mean value that is 99.73% confidence bounds.

As it was shown that majority of HFAs are developing at the tangential discontinuities (Schwartz et al., 2000) we calculate the normal vectors to the current sheet from each of 4 spacecraft data in the assumption of the tangential discontinuity, $\vec{n} = \frac{(\vec{B}_1 \times \vec{B}_2)}{|\vec{B}_1 \times \vec{B}_2|}$. These vectors are listed in Table 2. As the values obtained from C3 measurements that has very irregular structure are quite different from ones for 3 other spacecraft we have taken an "average" vector calculated from values for C1, C2, and C4 spacecraft.

Table 2. Normal vectors to the current sheet calculated from magnetic field measurements in assumption that the current sheet is a tangential discontinuity. Sign of the normal is chosen such that the vector is directed to the solar hemisphere (as proposed by Schwartz et al., 2010). For average component explanation see text.

| SC | $n_x$ | $n_y$ | $n_z$ |
|---|---|---|---|
| C1 | 0.726 | 0.209 | -0.656 |
| C2 | 0.823 | 0.417 | -0.384 |
| C3 | 0.209 | -0.414 | -0.886 |
| C4 | 0.858 | 0.476 | -0.196 |
| Average of C1, C2, C4 | 0.824 | 0.377 | -0.423 |
| Timing method | 0.457 | 0.889 | 0.023 |

Using the timing method for determination of discontinuity from 4-spacecraft data (Knetter at al., 2004) we came to normal vector direction listed in the last row of Table 2. It is quite different from one obtained with tangential discontinuity assumption.

The angle between the current sheet normal calculated with assumption of tangential discontinuity and sunward direction is $\theta_{X-CS} \sim 34.5°$. The same for the current sheet normal calculated by timing method is $\theta_{X-CS} \sim 62.8°$.

The normal to the bow shock at the location of 4 Cluster spacecraft was calculated with use of average bow shock model of Formisano et al., (1973). We need to allow for the shock displacement for the solar wind pressure $\rho V^2$ for current solar wind velocity and the number density that is factor of ~4.4 higher than average value. With $(\rho V^2)^{-6}$ dependence of magnetosphere's size on solar wind pressure the shock should be by factor of ~1.32 closer to the Earth than Cluster spacecraft. At this location magnetic field lines (assumed to be the straight lines) do not cross calculated shock dimensions. In order to ensure the magnetic field lines crossing the shock the scaling factor should be < 1.29 for $\boldsymbol{B_2}$ magnetic field and < 1.24 for both $\boldsymbol{B_1}$ and $\boldsymbol{B_2}$ magnetic field lines.

If we assume that Cluster spacecraft were close to the shock, we can scale the shock to Cluster position; the calculated $\theta_{Bn1}$ and $\theta_{Bn2}$ for magnetic fields at two sides of the current sheet are given in Table 3 with other relevant values. It means that independent of distance of the 4 Cluster spacecraft from the shock the magnetic field lines on both sides of the current sheet were connected to the quasi-perpendicular shock.

Table 3. Estimated $\theta_{Bn}$

| $N_{bs}$ | (0.872 ; - 0.210 ; - 0.441) |
|---|---|
| $B_1$ | (2.05, - 3.32, 1.38) |
| $B_1$ (normalized) | (0.495 ; - 0.802 ; 0.335) |
| $B_2$ | (2.16. - 2.80. 1.74) |
| $B_2$ (normalized) | (0.548 ; - 0.711 ; 0.441) |
| $\Theta_{bn1}$ | $65.5^0$ |
| $\Theta_{bn2}$ | $67.0^0$ |

## 2.3. Attribution of the current sheet

As the normal direction to the current sheet was not determined with sufficient confidence, it is not possible to verify if selected current sheet satisfies all criteria for the HFA (Schwartz, 2000). Existence of magnetic field enhancement at the edges of the current sheet, quasi-perpendicular bow shock geometry at least at one side of the current sheet and the criterion of high solar wind velocity facilitation (Facsko, 2009) to formation of HFA are supporting our definition. Wang et al., (2013a) have found that HFA can form in the case of quasi-perpendicular geometry at two sides of it.

This current sheet satisfies criterion of Foreshock Cavity (Sibeck et al., 2002) of small rotation of magnetic field vector across discontinuity, but does not satisfy condition of quasi-parallel shock connection on both sides of the current sheet. The last condition also makes this current sheet not attributed to the Foreshock Bubbles (Omidi et al., 2010).

With clear signs of multi-beam distributions this event can be attributed to the young HFAs (Zhang et al., 2010).

## 3. Upstream beam analysis

In order to analyze beam we separated velocity space in two domains (Figure 3). One domain was selected for calculation of velocity, temperature and number density of the solar wind. This domain is usually easy to separate as the solar wind flux is seen in well-defined direction. Other phase space was considered as the domain of beam. The number density, velocity and temperature of two populations were calculated as moments of the measured velocity distribution.

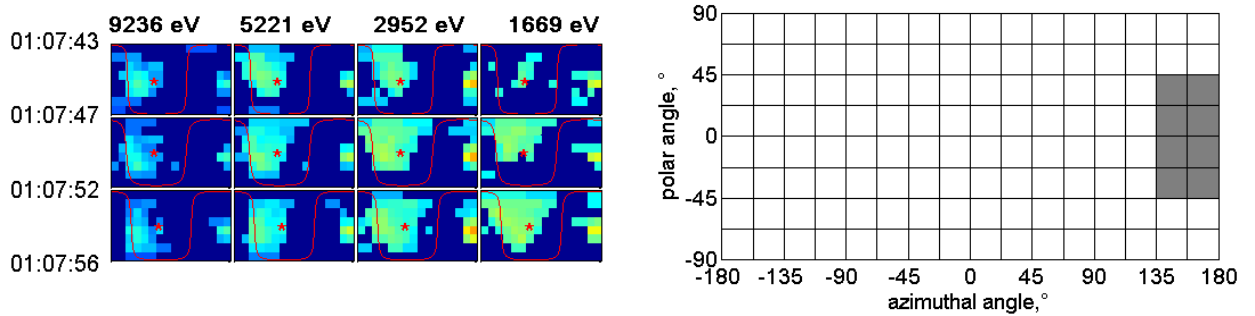

Figure 3. Velocity space domains selection. Left (a): several panels of (θ, φ) distributions of ion flux as measured by C1 spacecraft: energy at the top, time (hh:mm:ss) at the left. Red line shows magnetic equator and asterisk shows magnetic field direction. Bright spot at the middle right of each panel is the solar wind. Wide spot at the left is beam. Note nearly opposite directions of two beams. Right (b): complete (θ, φ) grid of velocity space scanned by ion analyzer in GSE coordinates. Grey sector indicates the solar wind beam location.

Event #3 is shown in Figure 4 in more detail. Upstream ion beam was observed at the leading side of HFA only. It showed velocity dispersion and had high temperature just within HFA (at 01:07:55 – 01:08:07 UT). Higher velocities were seen first, lower velocities were observed as the current sheet approached the spacecraft.

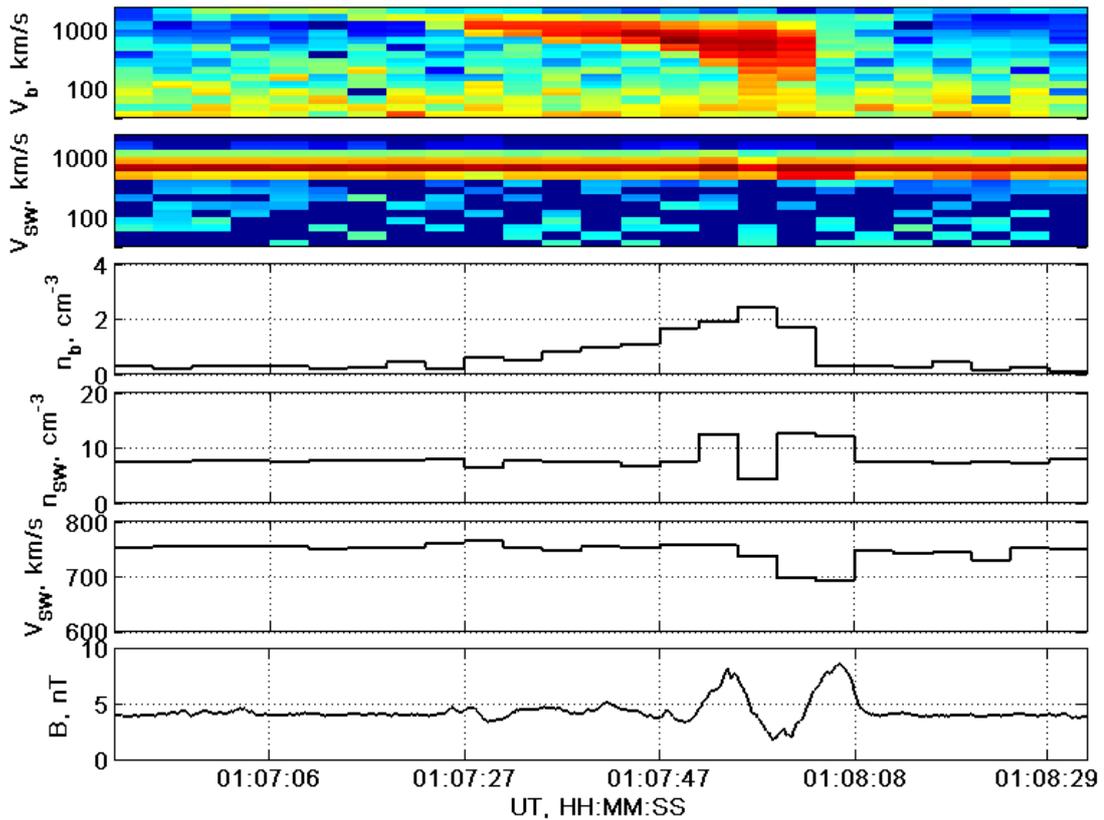

Figure 4. Event #3 as observed by C1 spacecraft. From up to bottom: velocity-time diagram of upstream beam, velocity-time diagram of the solar wind, upstream beam number density, solar wind number density, solar wind velocity, magnetic field magnitude. Plasma measurements temporal resolution is 4 s.

The main features of upstream beam are:
- It is observed on one side of the CS
- There is no gap between the beam and the CS
- Its average velocity decreases, the width of the spectrum and number density increase as CS approaches the spacecraft
- Number density of the beam just before HFA encounter (at 01:07:51-01:07:55 UT) was about 25% of the solar wind density
- Number density of the beam just within HFA (at 01:07:55 – 01:08:07 UT) was about 28% of the solar wind density
- Number density of the upstream beam reaches the maximum within the anomaly, being about 28% of the solar wind flux outside the CS
- The deceleration of the solar wind beam within HFA is ~ 50 km/s.

Figure 5 shows 2 series of velocity spectra as observed by C1 and C3 spacecraft. In spite of lower temporal resolution of C3 measurements there is significant similarity between variations of spectra observed on two spacecraft. It is worth to mention that C3 spacecraft is closer to the bow shock by ~ 5300 km.

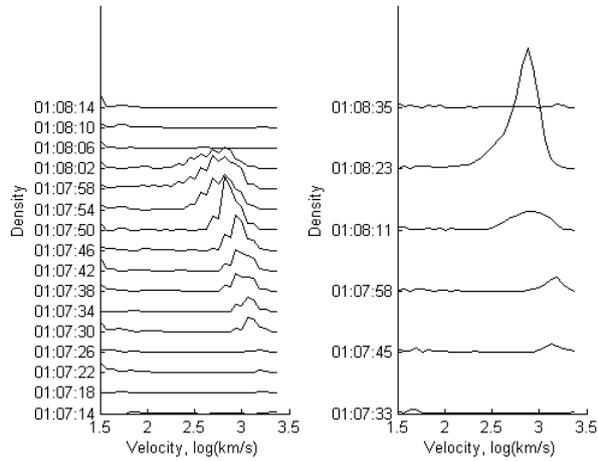

Figure 5. Left: sequence of ion velocity spactra as observed by C1 spacecraft, right: the same as observed on C3 spacecraft. Time is from bottom to top. Spectra are displaced along vertical scale for clarity, times for two spacecraft data are approximately adjusted with allowance for different temporal resolution of C1 measurements (4 sec) and one of C3 measurements (12 sec).

Ion velocity distributions are shown in composite picture in local magnetic coordinates. (Figure 6). Each rectangle is phase angle and pitch angle distribution of ions (assumed to be protons) phase space density for specific energy step from 170 eV to ~ 29900 eV (indicated at the top). Time as shown at the left is from top to bottom. Magnetic field magnitude, upstream beam density and solar wind beam density are shown at the left. Horizontal row is a snapshot of velocity distribution for the time inteval between times indicated on the left. The pitch and phase angle distribution of the current sheet changes as the CS approaches the C1 spacecraft. Solar wind is seen at energies from ~ 9200 eV to ~1670 eV at lower left of each panel (note different coordinates from Figure 3). Upstream beam is seen initially at upper part of panels as narrow beam at upper left of high-energy panels. Its width increases and finaly occupies the upper parts of the panels at the current sheet itself.

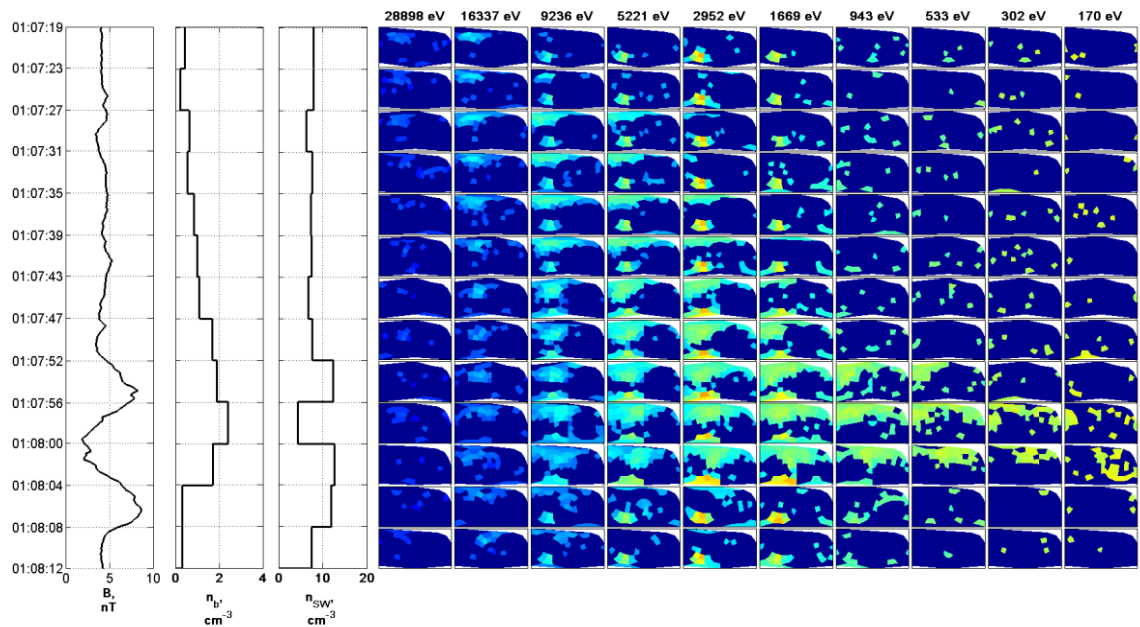

Figure 6. Upstream ions and solar wind velocity distributions as measured by C1 spacecraft. Each rectangular panel is pitch-angle (0° at the bottom and 180° at the top) - azimuth angle (0° at the left and 360° at the right) distribution for different proton energies (indicated at the top of each column). Time

(hh:mm:ss) is from top to bottom. At the left: magnetic field magnitude, upstream beam number density, solar wind number density. Temporal resolution of ion measurements is 4 s.

It is easy to see how energy width and angular span of upstream beam widens in time in all three dimensions of the phase space: energy, pitch-angle and phase angle. Finally it fills the gap to the solar wind beam within the current sheet (see the panel for energy 1669 eV in 08:00-08:04 time interval).

Similar behavior can be seen from measurements on C3 spacecraft (Figure 7) that is located by ~ 5300 km closer to the bow shock. It could be seen that the coupling of the upstream beam with the solar wind beam is significantly stronger that is manifested by wider bridge between two beams within the current sheet.

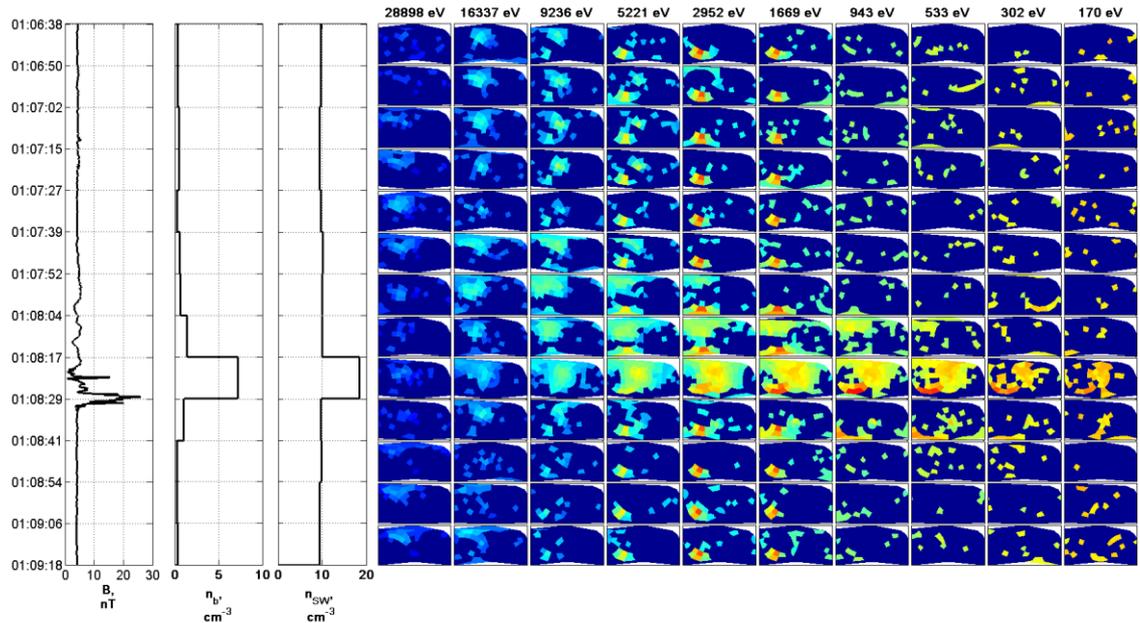

Figure 7. The same as at Fig. 6 but for C3 spacecraft. Note that temporal resolution of the measurements is lower than for C1 spacecraft, being 12 s instead of 4 s.

Figure 8 shows how the pitch-angle distributions of different energy ions of upstream beam changes while the current layer approaches C1 spacecraft. High energy ions are observed earlier than the lower ones. Ions of particular energy are initially observed at smaller pitch angles and they are seen at higher pitch angles later.

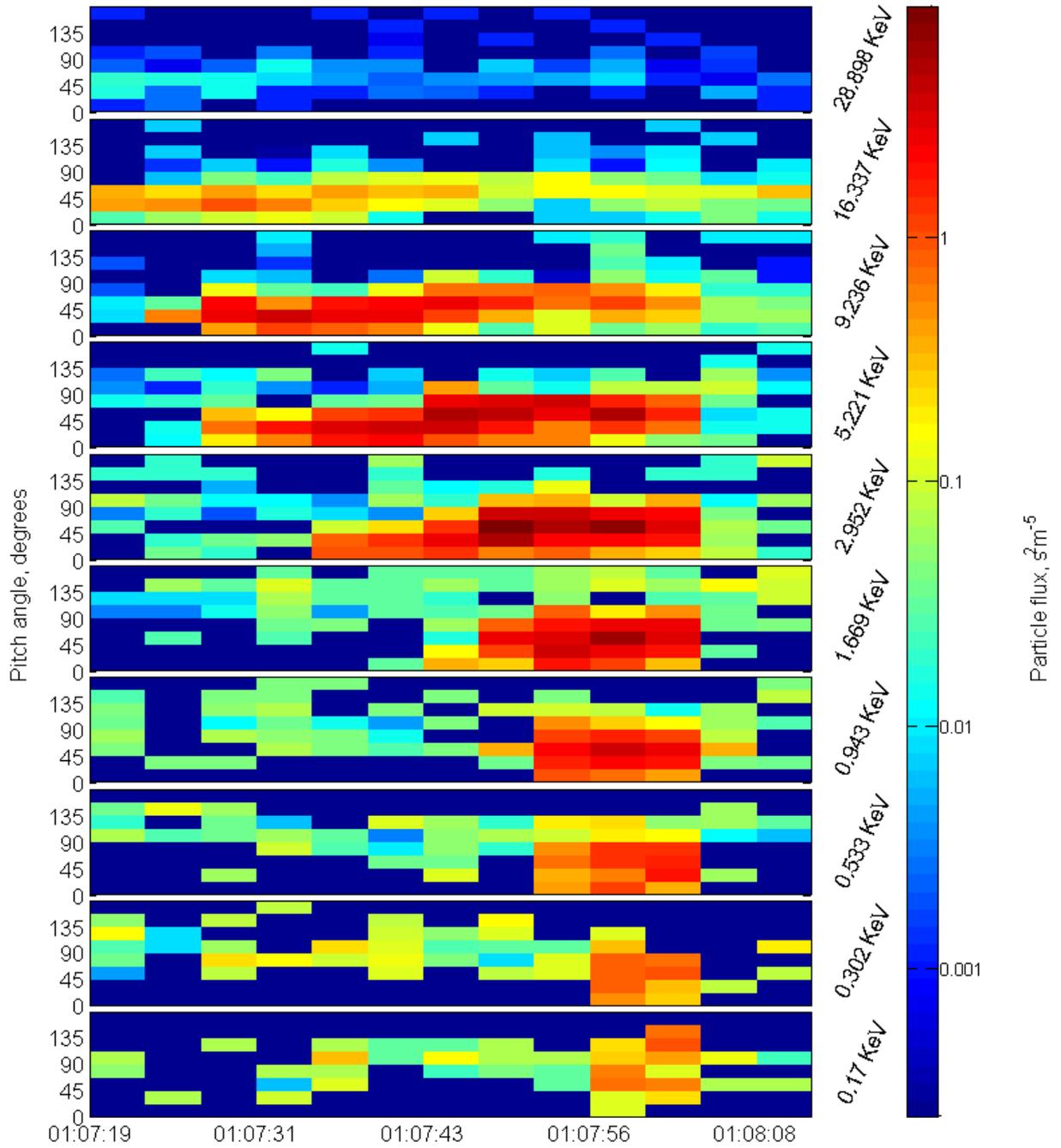

Figure 8. Variation of pitch-angle span of different energies (shown to the right of panels) versus time. Each panel is pitch-angle-time diagram for energy indicated at the right. Color scale indicates particle flux. Note that lower energies appear later in time.

The increase of the pitch angles from the first appearance of particular energy to the last registration within the current sheet suggests possibility that this increase is similar to all energies (except 0.30 keV and 0.17 keV) and may be scaled to the gyroradius ρ of these ions. Figure 9 is the verification of this possibility. The dependence of average pitch angle for particular time of first observation of each energy was scaled in time from the last observation in the current sheet by the factor of ratio $f = \rho_1/\rho_6$ where $\rho_1$ is gyroradius of ions with energy ~28.9 keV and $\rho_6$ is gyroradius of ions with energy ~1.67 keV (lower energies were observed only in 2 or 3 time intervals and do not allow to reveal the trend). This figure suggests that pitch-angles of ions of the beam had similar pattern of pitch-angle variations within the beam.

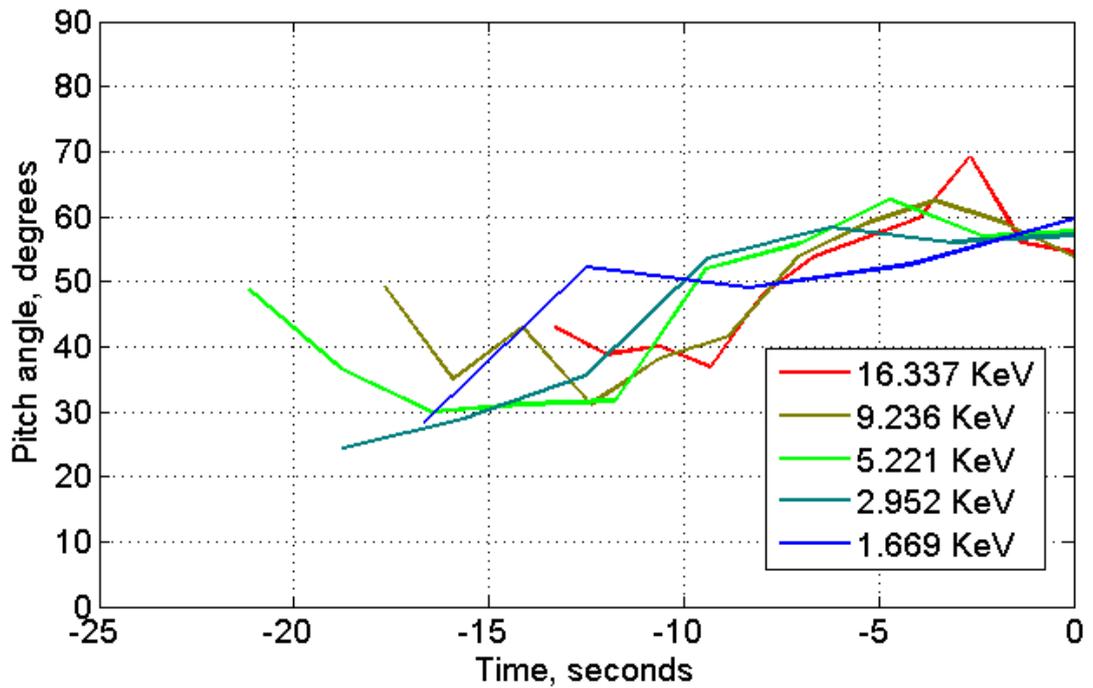

Figure 9. The beam energies with pitch-angle variations. Zero time is the current sheet crossing. Time scale for each energy was normalized by gyro-radius (see text).

Figure 10 shows the ion velocity versus the time of first observation of ions with this velocity in the beam. The least square fit by the straight line shows that the time interval between first registration of ions with specific velocity (and gyroradius) and the time of the current sheet observation is in approximate linear relation with velocity and, consequently, with ion gyroradius. There is no gap between upstream beam and the current sheet observations. The lowest energy beam ions of 0.30 keV and 170 eV are observed only within the current sheet itself.

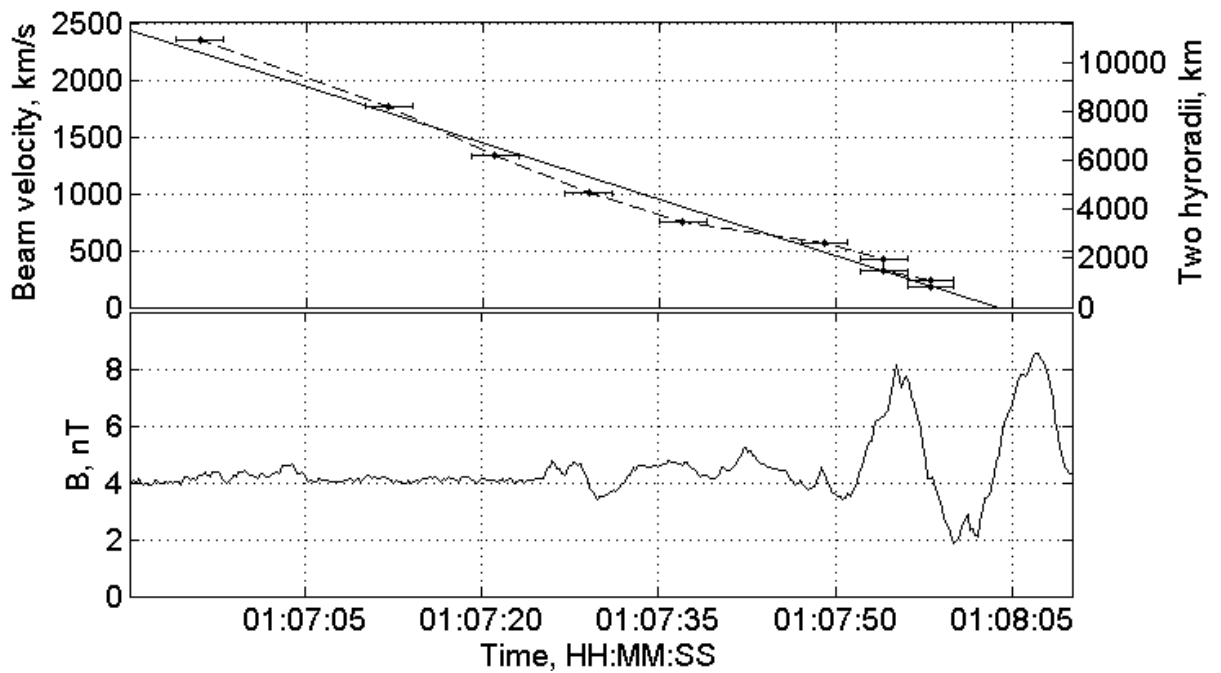

Figure 10. Above: velocity of the ions first observed on C1 spacecraft versus time (dots with time interval indicated). Mean square fit by the straight line is also shown. Below: profile of the magnetic field magnitude.

Distribution of magnetic and plasma pressures within and around current sheet are shown in Figure 11. It could be seen that both parallel and perpendicular ion pressures have maximum values within current sheet itself.

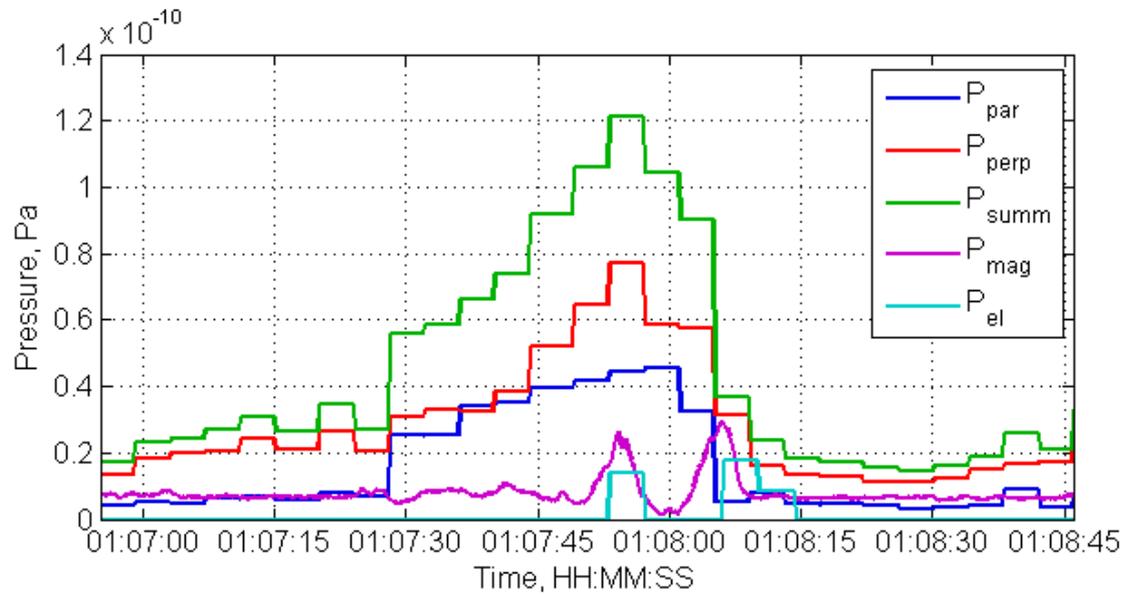

Figure 11. Magnetic field pressure (violet), ion parallel pressure (dark blue), ion perpendicular pressure (red), total ion pressure (green), and electron pressure (light blue, only three values are available).

The following properties of the upstream beam were found in this analysis:
- upstream ions were observed at the leading side of HFA only
- beam had velocity dispersion: high velocity ions observed first, lower velocities were observed as the current sheet approached the spacecraft
- the width of the ion spectrum increased as the current sheet approached the spacecraft
- the distance at which specific with velocity was nearly proportional to the ion gyroradius
- there was no gap between the beam and the current sheet
- lowest velocity ions of 0.17 keV and 0.30 keV were observed only within the current sheet
- density of ions increased towards the current sheet with maximum within current sheet
- the density of the ion beam was about 25-28% of undisturbed solar wind as observed on C1 spacecraft; it cannot be determined at C3 spacecraft located closer to the bow shock due to intermixing of velocity distribution beams
- the ion pressure has the maximum within current sheet
- the energy dissipation seen as interaction between upstream beam and the solar wind beam was stronger at the closer to the bow shock C3 spacecraft
- observed ion spectra just before the current sheet encounter and within it were quite similar to the ion spectrum in the magnetosheath behind the shock observed after ~ 70 minutes after HFA observation

Figure 12 shows four ion energy spectra: the spectrum of beam within current sheet, normalized sum of dispersed beam spectra recorded before CS crossing, spectrum of ions in front of nearest bow shock crossing, and integrated spectrum of magnetosheath ions with pitch-angles 0°-90° (in the hemisphere with velocities directed towards the shock front) just after bow shock crossing. The fittings of these spectra by convected Maxwellian distributions are also shown. Moments of distribution functions of these beams are given in Table 4. The properties of these ion spectra will be considered in section 4.

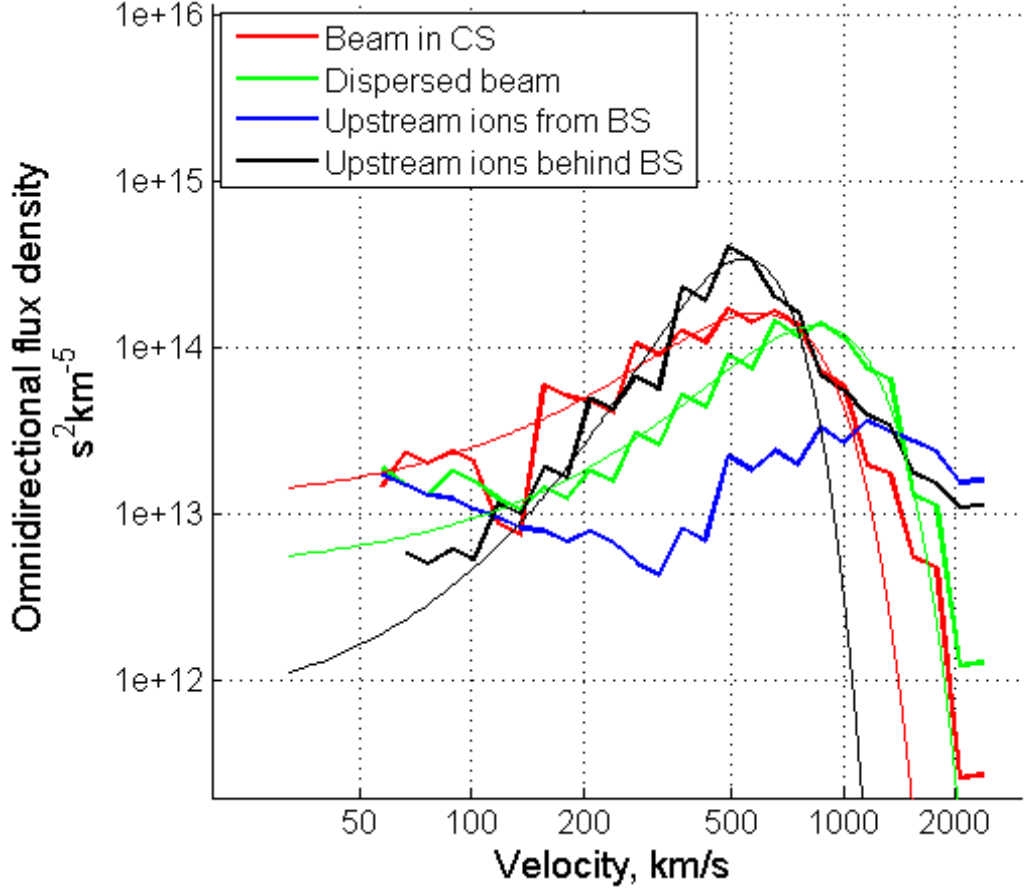

Figure 12. Comparison of ion spectral flux densities within current sheet (red), normalized cumulative sum of dispersed beam flux over all time interval before current sheet encounter (green), upstream ion beam in front of nearest shock crossing (blue), and magnetosheath ions with velocities directed to the shock front (black). Thin lines show fits by convected Maxwellian distributions. Please, note similarities in shape and flux magnitudes of red, green and black spectra and their differences from blue spectrum.

Table 4. Parameters of ion populations

|  | Dispersed beam at HFA | Beam within CS | Upstream ions behind BS |
|---|---|---|---|
| n, cm$^{-3}$ | 1.9 | 2.1 | 2.0 |
| $V_{par}$, km/s | 355.1 | 226.9 | 166.9 |
| $V_{perp}$, km/s | 423.8 | 344.8 | 508.8 |
| $V_{total}$, km/s | 552.9 | 412.3 | 535.5 |
| $T_{par}$, eV | 600.0 | 291 | 235 |
| $T_{perp}$, eV | 595.0 | 350 | 440 |

## 4. Discussion

Beam reflected from the bow shock was almost immediately recognized after discovery of the active current sheets as the agent responsible for the active current sheet development. Paschmann et al., (1988) mentioned as one of possible mechanisms for formation of anomalies at shocks the interaction of the bow shock with tangential discontinuities having a specific internal structure that results in the sudden and localized enhancements of bow shock reflection.

Burgess and Schwartz (1988) explained the mechanism interaction of the interplanetary current sheet by complete reflection of the incident flow at the intersection of the shock with transition to the downstream magnetic field strength value. This reflected ion component traverses the low-field region and occupies a region well upstream of the expected shock position based on the original shock speed.

Schwartz et al. (1988) suggested that these events are the direct result of the disruption and reformation of the bow shock by the passage of an interplanetary current sheet, most probably a tangential discontinuity.

Based on the analysis of ten coined Hot Diamagnetic Cavities Thomsen et al. (1988) found several events that were closely associated with intervals of dense, nearly specularly reflected ions. Wang et al., (2013b) found that stronger deflection of flow inside HFA are observed at larger distances from the bow shock subsolar point thus confirming important role of specularly reflected beam in formation of HFA.

Burgess (1989) studied the behavior of solar wind test particles specularly reflected at intersection of the current sheet with quasi-perpendicular shock (Figure 13). He concluded that the formation of active current sheets involves a large density of reflected ions at the contact point structures; current sheet and non-equilibrium behavior caused by a depletion of downstream gyrating ions may provide such over-reflection.

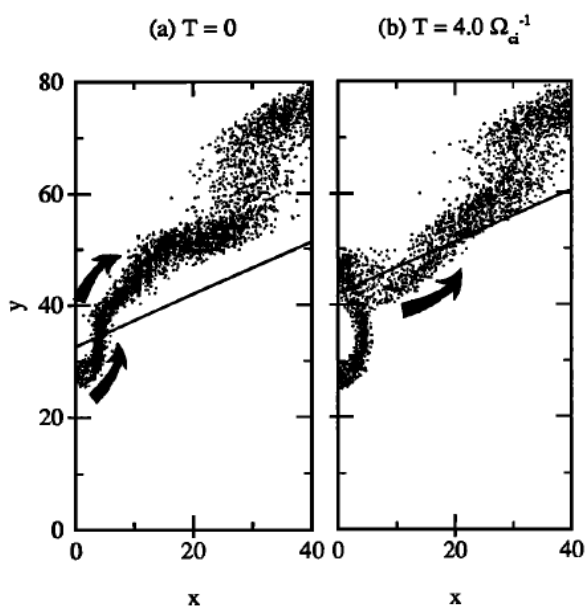

Figure 13. Test particles of ions reflected from strong shock (Burgess, 1989)

Thomas and Brecht (1988) studied instability created by reflected beam interaction with the solar wind and found that a high $\beta$ diamagnetic cavity of reduced averaged magnetic field and density is created by an ion beam of finite width by the electromagnetic ion beam instability.

In global hybrid simulation Omidi and Sibeck (2007) found that the HFA forms when TD reaches the location on the bow shock with quasi-parallel magnetic field orientation that allow ions to escape upstream and form HFAs.

Suprathermal ions upstream of shocks can sometimes originate from leakage of downstream ions. These processes were investigated by Edmiston et al. (1982) for different $\theta_{Bn}$ values. Results showed that leakage ions can be registered up to $\theta_{Bn} \sim 65°$ under certain conditions. Later Lyu and Kan (1990) presented simulation results on upstream ion leakage and reflection at quasiparallel shocks, obtained with one-dimensional hybrid-code simulation model. The results indicate that leakage ions can be predominant fraction at regions of $\theta_{Bn} \sim 10°$. Schwartz (1983) had also shown that magnetosheath particles leaked parallel to the shock normal reencounter the shock for $\theta_{Bn} > 65°$, but with insufficient energy to penetrate back into the downstream region.

Tjulin et al. (2009) analyzed velocity distribution of suprathermal dispersed ion beam observed at Hot Flow Anomaly. They were able to model pitch-angle and gyrophase distributions of this ion beam with HFA that was assumed to be the thin current layer. In the discussion Tjulin et al (2009) described relatively strict conditions for the adequacy of their model to observational conditions including the angle $\theta_{Bn}$ and close location of the spacecraft to the bow shock.

The beam we analyzed is in many respects similar to one discussed by Tjulin et al.(2009): this beam is suprathermal, dispersed and adjacent to the HFA like one we consider. However, it was decided not to use this model, as one needs to allow for the motion of the discontinuity. Also, comparable dimensions of the region occupied by the upstream beam and the HFA thickness require consideration of the current layer structure in the ion motion calculations.

The most interesting properties were found in the analyses: first ones are associated with the geometry and propagation of the ions relative to the current sheet, and second ones are characteristics of the beam: number density and energy.

1. Geometry and propagation:
- There is no gap between the energy-dispersed beam and the current sheet
- Ions of lowest energies, which naturally continue energy spectrum of upstream beam, are observed only within the current sheet
- Linear dependency of the time (or linear scale) between first observation of specific velocity of ions
- Similar pattern of pitch-angle variation of different energy ions within normalized by gyroradius distance from the current sheet
- Magnetic pressure and solar wind density are smaller in the central part of the current sheet than in surrounding space
2. Number density and energy characteristics:
- Beam number density increase towards the current sheet
- Highest number density is found in the current sheet itself
- The plasma pressure is higher in the current sheet than in surrounding flow
- Ion energy distribution of the beam is quite similar, but more heated, than in the magnetosheath

It is important that all these properties are seen at two spacecraft separated by a large distance. This also suggests that the beam analyzed in this paper is not a transient phenomenon, rather it is quasi-stationary structure associated with the current sheet. This is supported by the fact that C3 spacecraft was situated closer to the bow shock than C1 spacecraft but observed the beam with very similar characteristics later than C1, without signs of propagation effects in the beam properties. With ~ 5300 km distance spacecraft separation relative to the shock and the gyroradii of the ions in the beam in the range of 1000-5000 km the differences in the non-stationary reflected beam characteristics should be observable.

In order to analyze possible sources of the upstream beam one needs to compare observed beam within the current sheet itself and one of adjacent dispersed beam. Savoini et al. (2013) analyzed 4 different reflected ions population in the front of quasi-perpendicular shock with $\theta_{Bn}$ from 45° to 66°. Their model showed that the energy and pitch-angle distribution of reflected ions depends on the place within the shock structure. Our estimation of $\theta_{Bn}$ (Table 2) lies close to upper bound of this region, and steady flow of upstream ions observed in front of closest observed shock (blue curve in Fig. 12) is significantly more energetic than observed reflected beam at HFA under consideration. This does not support possibility that upstream beam which has properties of steady flow is due to reflection of ions from the shock front.

The linear velocity dispersion of the beam, smooth variations of beam characteristics with continuation of this trend to the beam in the current sheet with the maximum values of these characteristics within current sheet itself (note comparison of ion spectra in Figure 12 and table 4) suggest that the dispersed beam originate from the beam propagating upstream within the current sheet. Further evidence can be obtained from numerical simulation that is outside of the current analysis.

Not of less interest is the origin of the ion beam seen in the current sheet itself. Several properties of the beam propagating upstream within the current sheet (at 01:07:55 – 01:08:07 UT in Figure 4 and table 4) including number density and ion temperature suggest that the source of this beam can be magnetosheath. Comparison of ion spectrum within the current sheet of HFA and the spectrum of ions moving upstream within magnetosheath observed just behind the bow shock (Figure 12 and explanation of it in previous section) shows similarity of spectral shape and number density that supports proposed explanation of upstream beam origin.

The magnetic field and magnetic pressure within current sheet it has a minimum in the central region, and the solar wind flux is lower within the current sheet than in undisturbed solar wind (see the solar wind density variation in the Figure 4). This, and also magnetic field structure at the intersection of the current sheet and the bow shock may allow magnetosheath ions to stream within the current sheet and form the reflected beam. Observed properties of the beam suggest that its origin could be this low magnetic field region of the interplanetary current sheet at the intersection with the current flowing at the shock front. Region of two currents interaction could have very low magnetic part somewhat similar to one in the reconnection region. In this region ions become demagnetized and magnetosheath ions can escape through the interplanetary current sheet layer upstream to the solar wind flow. The properties of the reflected beam suggest that the ions observed by Cluster spacecraft on 22.02.2006 in front of the bow shock within HFA and in velocity-dispersed ion beam adjacent to the current sheet are the magnetosheath ions entering the current sheet within the intersection of it with the bow shock, and then channel within the current sheet upstream.

More detailed analysis requires numerical simulation to see if experimental evidence of magnetosheath ions to leak into the current sheet could be the reason for formation of the current sheet that was analyzed in this paper. Also, more cases need to be analyzed to see how widely this proposed process may be responsible for formation of reflected beam at the current sheets that cross the bow shock. Analysis of other events is in progress. Exact configuration of the resulting current system and the magnetic field can be understood by the computer modeling that we are planning as the next step.

## 5. Conclusion

Analysis the reflected beam properties within the current sheet of the Hot Flow Anomaly and associated dispersed beam at the leading side of HFA shows several geometry and physical properties that suggests the propagation of the upstream beam inside the current sheet. The possible explanation of the HFA with upstream beam observed by Cluster spacecraft at 01:09 UT on 22.02.2006 is that the beam forms due to escape of magnetosheath ions entering in the region of the interplanetary current sheet crossing of the bow shock. They can subsequently propagate upstream within the interplanetary current sheet. The ions with gyroradius larger than the current structure can form the dispersed beam within the region adjacent to the current sheet.

## 6. Acknowledgement

Authors are grateful to: H.Reme, I.Dandouras, J.-A.Sauvaud and E.Penou for discussions while working with plasma data from Cluster databases. One of authors is grateful to Steven

Schwartz for discussion of our poster at the 2015 Fall AGU meeting, specifically about the solar wind depression.

# 7. References

Burgess, D., On the Effect of a Tangential Discontinuity on Ions Specularly Reflected at an Oblique Shock, J. Geophys. Res., Vol. 94, No. A1, Pages 472-478, January 1, 1989

Burgess, D., and S. J. Schwartz, Colliding plasma structures: Current sheet and perpendicular shock, J. Geophys. Res., 93, 11,327-1!,340, 1988

Facsko, G., et al., 2009, A global study of hot flow anomalies using Cluster multi-spacecraft measurements, Annales Geophysicae, 05/2009; 27(5). DOI: 10.5194/angeo-27-2057-2009

Formisano, V., Orientation and shape of the Earth′s bow shock in three dimensions, Planetary and Space Science, VOL.27, p. 1151-1161, 1979.

Gary S., Electromagnetic ion/ion instabilities and their consequences in space plasmas - A review, Space Science Reviews, vol. 56, p. 373-415, 1991.

Knetter, T., Neubauer, F.M., Horbury, T. and Balogh, A. (2004). Four- point discontinuity observations using Cluster magnetic field data: A statistical survey. J. Geophys..Res., 109: doi: 10.1029/2003JA010099. issn: 0148-0227.

Omidi, N. and G.Sibeck, Formation of hot flow anomales and solitary shockes, J. Geophys. Rees., Vol. 112, A01203, doi: 10.1029/2006JA011663, 2007.

Omidi, N., J. P. Eastwood, and D. G. Sibeck (2010), Foreshock bubbles and their global magnetospheric impacts, J. Geophys. Res., 115, A06204, doi:10.1029/2009JA014828.

Paschmann, G., G. Haerendel, N . Sckopke, E. Mobius, H. Liihr, C. W. Carlson, Three-Dimensional Plasma Structures with Anomalous Flow Directions near the Earth's Bow Shock, J. Geophys. Res., Vol. 93, No. A10, Pages 11,279-11,294, October 1, 1988

Savoini, P., B.Lembege, and J.Steinlet, On the origin of the quasi-perpendicular ion foreshock: Full-particle simulation, J.Geophys. Res., Vol. 118, 1132-1145, 2013, doi: 10.1002/jgra.50158.

Schwartz S.J., Chaloner C.P., Hall D.S., Christiansen P.J., Johnstones A.D., An active current sheet in the solar wind Nature (ISSN 0028-0836), vol. 318, Nov. 21, p. 269-271, 1985.

Schwartz S.J., Paschmann G., Sckopke N., Bauer T., Dunlop M., Fazakerley A., Thomsen M., Conditions for the formation of hot flow anomalies at Earth's bow shock, J. Geophys. Res.,105, 12639-12650, 2000.

Sibeck, D. G., T.-D. Phan, R. Lin, R. P. Lepping, and A. Szabo, Wind observations of foreshock cavities: A case study, J. Geophys. Res., 107(A10), 1271, doi:10.1029/2001JA007539, 2002.

Steven J . Schwartz, Ramona L . Kessel, Cassandra C. Brown, Les J. C. Woolliscroft , Malcolm W . Dunlop, Charles J . Farrugia, and David S. Hall, Active Current Sheets near the Earth's Bow Shock, Journal of Geophysical Research, Vol. 93, No. A10, Pages 11,295-11,310, October 1, 1988

Steven J . Schwartz, Ramona L . Kessel, Cassandra C. Brown, Les J. C. Woolliscroft , Malcolm W . Dunlop, Charles J . Farrugia, and David S. Hall, Active Current Sheets near the Earth's Bow Shock, Journal of Geophysical Research, Vol. 93, No. A10, Pages 11,295-11,310, October 1, 1988

Thomas, V. A. and S.H. Brecht, Evolution of Diamagnetic Cavities in the Solar Wind, Journal of Geophysical Research, Vol. 93, No. A10, Pages 11,341-11,353, October 1, 1988

Thomsen, M.F., J.T.Gosling, S. J. Bame, K.B.Quest, C.T.Russell, S. A. Fuselier, On the origin of Hot Diamagnetic Cavities Near the Earth's Bow Shock, J.Geophys. Res., Vol. 93, No. A10, Pages 11,311-11,325, 1988.

Tjulin, A., E. A. Lucek, and I. Dandouras, Observations and modeling of particle dispersion signatures at a hot flow anomaly, J. Geophys. Res., VOL. 114, A06208, doi:10.1029/2009JA014065, 2009

Turner, D. L., N. Omidi, D. G. Sibeck, and V. Angelopoulos (2013), First observations of foreshock bubbles upstream of Earth's bow shock: Characteristics and comparisons to HFAs, J. Geophys. Res. Space Physics, 118, 1552–1570, doi:10.1002/jgra.50198.

Wang S., Q. Zong Q.-G., Zhang, H., Cases and statistical study on Hot Flow Anomalies with Cluster spacecraft data. Sci China Tech Sci, 2012, 55: 1402–1418, doi: 10.1007/s11431-012-4767-z

Wang S., Q. Zong and H. Zhang , Cluster observations of hot flow anomalies with large flow deflections: 1. Velocity deflections, J. Geophys. Res.; Space Physics, 118, 2013a, doi:10.1029/2012JA017833.

Wang, S., Q. Zong, and H. Zhang, Cluster observations of hot flow anomalies with large flow deflections: 2. Bow shock geometry at HFA edges, J. Geophys. Res. Space Physics, 118, 418–433, 2013b, doi:10.1029/2012JA018204.

Zhang, H., D. G. Sibeck, Q.-G. Zong, S. P. Gary, J. P. McFadden, D. Larson, K.-H. Glassmeier, and V. Angelopoulos, Time History of Events and Macroscale Interactions during Substorms observations of a series of hot flow anomaly events J. Geophys. Res., VOL. 115, A12235, doi:10.1029/2009JA015180, 2010